\documentclass[letterpaper,11pt]{article}
\pdfoutput=1
\usepackage{jheppub}
\usepackage{subfigure}
\usepackage{amsmath}
\usepackage{multirow}
\usepackage{tablefootnote}
\usepackage{comment}
\usepackage{hyperref}
\usepackage{epstopdf}

\newcommand{\beq}{\begin{equation}}
\newcommand{\eeq}{\end{equation}}
\newcommand{\bea}{\begin{eqnarray}}
\newcommand{\eea}{\end{eqnarray}}

\def\m1{M_1}
\def\m2{M_2}
\def\m3{M_3}

\def\br{\rm Br}

\def\ch10{\tilde \chi^0_1}

\def\mh{m_{h}}

\def\gev{\,{\rm GeV}}
\def\mev{\,{\rm MeV}}

\def\to{\rightarrow}

\newcommand{\lsim}{\mathrel{\mathop{\kern 0pt \rlap
  {\raise.2ex\hbox{$<$}}}
  \lower.9ex\hbox{\kern-.190em $\sim$}}}
\newcommand{\gsim}{\mathrel{\mathop{\kern 0pt \rlap
  {\raise.2ex\hbox{$>$}}}
  \lower.9ex\hbox{\kern-.190em $\sim$}}}

\definecolor{pink}{RGB}{255,105,180}

\def\fbi{\,{\rm fb}^{-1}}
\def\abi{\,{\rm ab}^{-1}}

\newcommand{\mm}{\ensuremath{\mu^+\mu^-}}
\newcommand{\ee}{\ensuremath{e^+e^-}}

\newcommand{\Tao}[1]{\textcolor{magenta} {Tao: #1}}
\newcommand{\ZL}[1]{\textcolor{red} {Zhen: #1}}
\newcommand{\Edit}[1]{\textcolor{black} {#1}}

\title{ISR effects for resonant Higgs production at future lepton colliders}

\author[a]{Mario Greco,}
\author[b,c]{Tao Han}
\author[d]{and Zhen Liu}

\affiliation[a]{Dipartimento di Matematica e Fisica, Universita' di Roma Tre, 
INFN, Sezione di Roma Tre,\\
Via della Vasca Navale 84, I-00146 Rome, Italy}
\affiliation[b]{Pittsburgh Particle physics, Astrophysics, and Cosmology Center, \\
Department of Physics and Astronomy, University of Pittsburgh, \\
3941 O'Hara St., Pittsburgh, PA 15260, U.S.A.}
\affiliation[c]{Department of Physics, Tsinghua University, Beijing,  100086, 
and Collaborative Innovation Center of Quantum Matter, Beijing, China}
\affiliation[d]{Theoretical Physics Department, Fermi National Accelerator Laboratory, Batavia, IL, 60510}

\emailAdd{mario.greco@roma3.infn.it}
\emailAdd{than@pitt.edu}
\emailAdd{zliu2@fnal.gov}

\abstract{We study the effects of the initial state radiation on the $s$-channel Higgs boson resonant production at $\mm$ and $\ee$ colliders by convoluting with the beam energy spread profile of the collider and the Breit-Wigner resonance profile of the signal. We assess their impact
on both the Higgs signal and SM backgrounds for the leading decay channels 
$h\to b\bar b,~WW^*$. 
Our study improves the existing analyses of the proposed future resonant Higgs factories and provides further guidance for the accelerator designs with respect to the physical goals. }


\preprint{
\begin{flushright}
RM3-TH/16-6\\
PITT PACC-1601\\
FERMILAB-PUB-16-261-T
\end{flushright}
}

\begin{document}

\maketitle
\flushbottom

\section{Introduction}

The Higgs boson discovery at the LHC in 2012~\cite{Aad:2012tfa,Chatrchyan:2012xdj} has opened a new era of particle physics: It is the first elementary scalar particle ever observed in Nature and its properties thus need to be thoroughly scrutinized. 
Future lepton collider Higgs factories~\cite{Barger:1995hr,Barger:1996jm,Han:2012rb,Alexahin:2013vla,Alexahin:2013ojp,Rubbia:2013iya,Baer:2013cma,Gomez-Ceballos:2013zzn,CEPCPreCDR,CEPCPreCDRvolumn2} are proposed to study the Higgs boson properties to great accuracies because of the much more favorable experimental environment than that at hadron colliders~\cite{Dawson:2013bba}. Amongst many candidates of Higgs factories, the possibility of $s$-channel resonant production is especially important. The muon collider Higgs factory could produce the Higgs boson in the $s$-channel and perform an energy scan to map out the Higgs resonance line shape at tens of MeV level \cite{Barger:1995hr,Barger:1996jm,Han:2012rb}. This approach  would provide the most direct measurement of the Higgs boson total width and the Yukawa coupling to  muons. The clean environment of the lepton colliders with a large number of Higgs bosons produced also enables precision measurements for many exclusive decays of the Higgs boson. More recently the possibility of an ultra high luminosity electron-positron collider for the Higgs resonant production has been proposed \cite{FCCeeHiggs}, providing a  possible opportunity to observe the Higgs signal and thus the determination of the Yukawa coupling to electrons $-$ so far the only conceivable measurement of the Higgs coupling to the first generation of fermions~\cite{Altmannshofer:2015qra}.

Due to the narrow width of the Higgs boson, about $4.07$~MeV~\cite{Dittmaier:2012vm} as predicted by the Standard Model (SM),  it would be extremely demanding for the collider energy resolution to reach a similar value in order to adequately study the physical width. This has been quantified in the literature by convoluting the Breit-Wigner resonance for the Higgs signal and the Gaussian distribution for the profile of beam energy spread (BES) \cite{Barger:1995hr,Han:2012rb}. 
It is also known that, the Initial State Radiation (ISR) of the QED effect would degrade the peak luminosity of a lepton collider~\cite{Barger:1996jm}. 
The impact on a muon collider has been recently emphasized \cite{Greco:2015yra,Jadach:2015cwa}, and the effects would be notably stronger for an $e^+e^-$ collider because of a lighter electron mass. 
In this work, we study all the effects coherently for a few representative choices of the BES and  different approximations for the ISR. 
%
We assess their impact in different scenarios on both the Higgs signal and SM background. Our study improves the existing analyses of the proposed future resonant Higgs factories and provides further guidance for the target accelerator designs with respect to the physical goals.

The work is organized as follows. In Section 2 we present the formulation and parameterization of the BES and ISR effects. In Section 3 we quantify their effects on the Higgs boson signal and the SM backgrounds on a muon collider, and study the observability for the Higgs signal at an $e^+e^-$ collider. We conclude in Section 4. Some analytical formulas adopted in our calculations for the ISR are listed in Appendix~\ref{app:ISR}.

\section{BES and ISR Parameterization in Resonant Higgs Boson Production}


\subsection{BES Parameterization}

As studied to a great detail in the literature \cite{Barger:1995hr,Barger:1996jm,Han:2012rb}, the muon collider energy resolution is critically important to study the Higgs width and interactions due to the very narrow width of the Higgs boson. The observable cross section is given by the convolution of the energy distribution delivered by the collider. We assume that the lepton collider c.m.~energy ($\sqrt s$) has a flux $L$ distribution
$$ { dL(\sqrt s) \over d\sqrt{\hat s} }
= {\frac 1 {\sqrt{2\pi \Delta}} } \exp[\frac {-( \sqrt{\hat s} - \sqrt s)^2} {2\Delta^2}] ,
$$
with a Gaussian energy spread $\Delta  = R \sqrt {s}/\sqrt{2}$, where $R$ is the percentage beam energy resolution, then the effective cross section is
\bea
\label{eq:convol}
&& \sigma_{\rm eff}(s) = \int d \sqrt{\hat s}\  \frac {dL(\sqrt s)} {d\sqrt{\hat s}} \  \sigma(\ell^+\ell^- \to h \to X)(\hat s) \\
&& \propto  \left\{
\begin{array}{ll}
\Gamma_h^{2} B / [( s-m^2_h)^2 + \Gamma_h^2 m_h^2]
\quad~   (\Delta  \ll \Gamma_{h}), & \\
B \exp[{ \frac {-( m_h - \sqrt s)^2 } {2\Delta^2} }]
(\frac {\Gamma_h}  {\Delta}) / m^{2}_h
  \quad  (\Delta \gg \Gamma_{h}).  &
\end{array}
\right.
\nonumber
\eea
The interaction strength $B$ is proportional to the Higgs coupling squared and governs the  overall normalization for the Higgs production rate.
For  $\Delta  \ll \Gamma_{h}$, the line shape of a Breit-Wigner resonance can be mapped out by scanning over the energy $\sqrt s$ as given in the first equation.
For $\Delta  \gg \Gamma_{h}$ on the other hand, the physical line shape is smeared out by the Gaussian distribution of the beam energy spread and the signal rate will be determined by the overlap of the Breit-Wigner and the luminosity distributions, as seen in the second equation above.
%
Our results for the Higgs line-shape  will be discussed in detail in the next section.  In addition to the Higgs signal, an important issue of phenomenological interest is the question of the expected background in the various Higgs decay channels. That is mainly related to the tail of the $Z$-boson produced in the lepton annihilation. This issue has already been a subject of study in Ref.~\cite{Han:2012rb}, but has to be reviewed in the light of the corrections introduced by the radiative effects.
Again those corrections will be calculated using the theoretical approach discussed above. Then the results for the Signal/Background ratio for various final state configurations will be also discussed in the next section. 

\subsection{ISR Parameterization}

Multiple soft photon radiation in the initial state (ISR) is an important effect in high energy lepton collisions~\cite{Yennie:1961ad}. In particular, when a narrow resonance is produced in an $s$-channel annihilation, the ISR effect becomes more significant. The first prominent example of such effects was the historical observation of $J/\psi$ production in $e^+e^-$ annihilation~\cite{Augustin:1974xw}. 
The origin of the ISR effect is well-known and was earlier discussed in great detail near the $J/\psi$ peak \cite{Greco:1975rm}, and later for the case of the $Z$-boson production~\cite{Greco:1980mh}. Qualitatively, a modification factor to the lowest order cross section can be expressed by
$$\kappa \propto ({\Gamma\over M})^{ {4\alpha\over \pi} \log({\frac {\sqrt{\hat s}}  m})}$$
where $M$ and $\Gamma$ are the mass and width of the $s$-channel resonance, 
$\sqrt{\hat  s}$ is the c.m.~energy in the partonic collision, and $m$ is the beam lepton mass. Physically this implies that the width provides a natural cut-off in damping the energy loss for radiation in the initial state. Very precise calculation techniques  for these QED effects have been developed for LEP experiments, where in addition to multi-photon radiation finite corrections have been added, by including, at the least, up to two-loop effects~\cite{Ellis:1986jba,Ellis:1986ja}.
In the case of muon colliders, in particular for Higgs boson production studies,  those effects were not emphasized sufficiently in the past, and only recently their importance has been pointed out~\cite{Greco:2015yra,Jadach:2015cwa} for the experimental study of the Higgs line-shape as well as for the machine design of the initial BES. In particular the estimates of the reduction factors of the Higgs production cross sections, of order of 50\% or more, depending upon the machine energy spread, given in Ref.~\cite{Greco:2015yra},  have been confirmed in Ref.~\cite{Jadach:2015cwa}, with the Higgs line-shape explicitly shown.


We will make use of the general formalism of the electron (muon) structure functions, first introduced in Ref.~\cite{Kuraev:1985hb}, and later improved for LEP experiments, which is well suited for the numerical calculations of the various distributions of phenomenological interest. 
\Edit{Our goal is to produce integrated cross sections to an accuracy of $O(1\%)$ which could be used as a reference in current studies of lepton Higgs factories. For the sake of completeness,
we will compare various levels of the approximation which can be found in the literature for the lepton structure functions.} As a first calculation technique we will use the approach of Ref.~\cite{Nicrosini:1986sm}, where in addition to the exponentiated effect from multi-photon radiation, finite terms have been included up to the second order. The results will be compared with the approach discussed in Ref.~\cite{Jadach:2000ir} \Edit{--- to various levels of approximation ---} and explicitly adopted in Ref.~\cite{Jadach:2015cwa} for the Higgs line-shape.

The initial state radiation (ISR) effect collectively can be expressed with different levels of sophistication. We define the probability distribution function $f^{\rm ISR}_{\ell\ell}(x)$ for the hard collision energy $x \sqrt {\hat s}$, and hard collision cross section
\beq
\sigma(\ell^+\ell^-\to h \to X)(\hat s)=\int d x~f_{\ell\ell}^{\rm ISR}(x;\hat s)\hat \sigma(\ell^+\ell^-\to h \to X)(x^2\hat s),
\eeq
where $x$ is the fraction of the c.~m.~energy at the hard collision with respect to the beam energy before the collision. 
We list several commonly used analytical formulas for the ISR parameterization under different approximations in Appendix~\ref{app:ISR}.

In particular, \Edit{for completeness and the convenience of the readers, we summarize here the various approximations \Edit{in the literature}. In Ref.~\cite{Kuraev:1985hb} only  some $O(\alpha)$ terms are included in addition to the exponentiated soft radiation term. In Ref.~\cite{Jadach:2000ir}  the various approximations  contain: (a) the soft exponentiated term only; (b) adds to (a) the full $O(\alpha)$ terms; and (c)  adds to (b) the relevant $O(\alpha^2)$ terms. Finally Ref.~\cite{Nicrosini:1986sm} contains the full exponentiated term and the complete $O(\alpha)$  and $O(\alpha^2)$ terms. The explicit expressions of various approximations are provided in Appendix~\ref{app:ISR}. }
%

%
\begin{figure}[t]
\centering
\subfigure{
\centering
\includegraphics[width=0.485\textwidth]{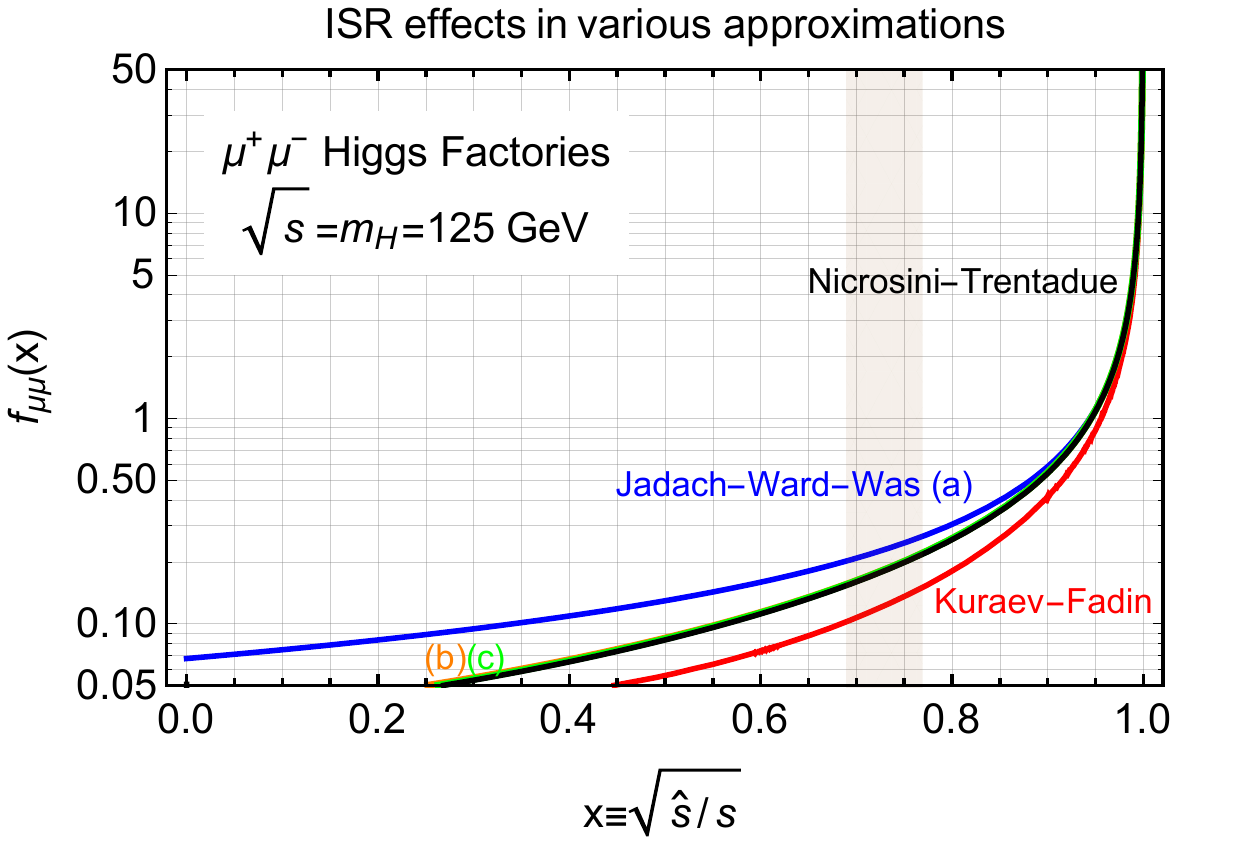}}
\subfigure{
\centering
\includegraphics[width=0.485\textwidth]{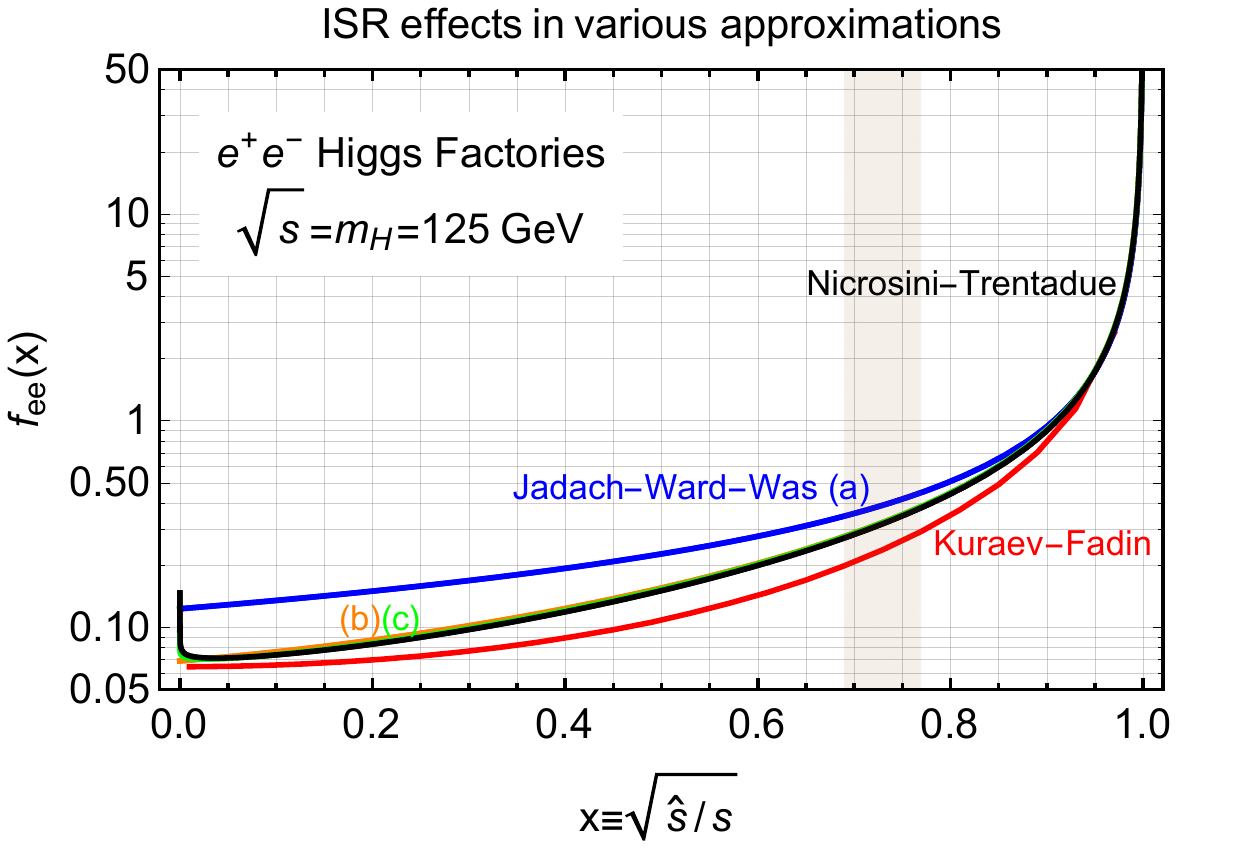}}
\caption[]{The beam energy distribution as a function of the energy fraction $x$ from ISR effects in various approximations, for $\mu^+\mu^-$  (left panel) and  $e^+e^-$ (right panel) initial states for 125 GeV center of mass energy. 
The shaded brown bands correspond to the collision energy near the $Z$-boson mass in the $\pm2\Gamma_Z$ window. 
}\label{fig:ISRpdf} 
\end{figure}

In Fig.~\ref{fig:ISRpdf}, we show those energy distributions versus the energy fraction $x$ with the ISR effects in various approximations, for $\mu^+\mu^-$  (left panel) and  $e^+e^-$ (right panel) initial beams for a c.~m.~energy $\sqrt s =125$ GeV. 
%
We can see that the widely used Kuraev-Fadin~\cite{Kuraev:1985hb} approximation (lower red curves) is more steep in falling comparing to other improved calculations. 
\Edit{The approach of} Jadach-Ward-Was~\cite{Jadach:2000ir} improves the approximations at different orders and complexities as listed in Appendix \ref{app:ISR}. We see that their benchmark choice (a) leads to a much larger radiation tail (upper blue curves). On the other hand, 
the more sophisticated approximations of Ref.~\cite{Jadach:2000ir} (b) and (c) (middle orange and green curves) agree \Edit{much better} with earlier works by Nicrosini and Trentadue~\cite{Nicrosini:1986sm} (shown as the black curves). This comparison signifies the importance of the proper treatment in evaluating the ISR effects. 
\Edit{Henceforth, we will restrict to the approaches of Ref.~\cite{Nicrosini:1986sm} and the choices (b) and (c) of Ref.~\cite{Jadach:2000ir}, which will lead us to our final calculated cross sections with the estimate of the theoretical  accuracy. In particular we have found that  at the level of the structure functions $f_{\ell\ell}^{\rm ISR}(x;\hat s)$ the relative difference between the formalism (b) by Jadach-Ward-Was~\cite{Jadach:2000ir}  and that of Nicrosini-Trentadue~\cite{Nicrosini:1986sm} is at 1$\sim$2\% level and 4$\sim$5\% level for large and small values of x, respectively.  However the relative difference between the calculated Higgs cross sections is smaller  because in the convolution only high-$x$ matter. }
The $\mu^+\mu^-$ case has \Edit{of course a} better agreement comparing to the $e^+e^-$ case because of the \Edit{smaller radiative effects} of muons.
\Edit{In particular, in the muon case, we have checked that the relative difference between the formalism (b) by Jadach-Ward-Was~\cite{Jadach:2000ir}  and that of Nicrosini-Trentadue~\cite{Nicrosini:1986sm} for the final cross sections is within 1\%, which is the estimate of the theoretical accuracy of our results.  On the other hand the cross sections difference between the formalisms (b)  and (c) by Jadach-Ward-Was~\cite{Jadach:2000ir} is O(0.1\%). }
As we are performing a convolution of the ISR effect over BES effect and then over the Breit-Wigner profile for a scan, the computational accessibility is important here. 
\Edit{We have found that the evaluation time for the structure functions alone in the formalism of Nicrosini-Trentadue~\cite{Nicrosini:1986sm} is about 5 times larger than the  formalism (b) by Jadach-Ward-Was~\cite{Jadach:2000ir}, at any value of $x$.}
Consequently, we choose formalism (b) as a balanced formalism between speed and accuracy.
\Edit{Our results, as stated above, will have a theoretical accuracy of O(1\%).}

As a consequence of the ISR, a very significant phenomenon  is the ``radiative return'' to a lower mass resonance. Despite the beam collision energy is above a resonance mass, after ISR radiation, the hard collision center of mass energy ``returns'' to the resonance mass and hit the Breit-Wigner enhancement again. This mechanism can be utilized to effectively producing lighter resonances without scanning the beam energy~\cite{Chakrabarty:2014pja,Karliner:2015tga}. 
In Fig.~\ref{fig:ISRpdf}, we shade the region in brown color for the  $x$ values corresponding to the $\pm 2\Gamma_Z$ window near $Z$-boson mass for a 125 GeV lepton collider. The rate in this window predicts the amount of ``radiative return'' $Z$ bosons produced, which constitutes a large background for Higgs studies. 
Once again, we can see that different parameterizations of the ISR effects yield significantly different amount of ``radiative return'' $Z$ production rate. 


\section{Numerical studies on the ISR and beam effects}

The ISR effects, as discussed in details in previous sections, are very important and inevitable at future lepton collider resonant Higgs factories. The ISR effects need to be convoluted with the finite BES as expressed in Eq.~(\ref{eq:convol}). We evaluate numerically their importance in the Higgs boson property measurements in this section.

\subsection{The case for the muon collider}
The muon collider Higgs factory features a line-shape scan of the Higgs boson, enables a simultaneous measurement of the Higgs boson mass, width and muon Yukawa at unprecedented precision~\cite{Barger:1995hr,Barger:1996jm,Han:2012rb}. The inclusion of the ISR effects make the prediction more robust. 

\begin{table}[tb]
\centering
\begin{tabular}{|c|c|c|c|c|}
  \hline
$~{}~~~~~~~~\sigma$(BW) ${}~~~~~~$ & ISR alone & { R (\%)} &  BES alone &  BES+ISR  \\ \hline
\multirow{2}{*}{$\mm$:~71~{\bf pb}}\hfill & \multirow{2}{*}{37} & $0.01$ & $17$ & $10$\\ \cline{3-5}
 & & $0.003$ & $41$ & $22$ \\ \hline
 \hline
\multirow{2}{*}{$\ee$:~1.7~{\bf fb}} & \multirow{2}{*}{0.50} & $0.04$ & $0.12$ & $0.048$\\ \cline{3-5}
 & & $0.01$ & $0.41$ & $0.15$ \\ \hline
\end{tabular}
\caption{Effective cross sections in $\mm$ (upper panel) collision in units of pb and $\ee$ (lower panel) collision in units of fb at the resonance $\sqrt {s}=\mh=125~\gev$, with Breit-Wigner resonance profile alone, with ISR alone (Jadach-Ward-Was (b)), with BES alone for two choices of beam energy resolutions,
and both the BES and ISR effects included. 
}
\label{tab:muc_onpeak}
\end{table}

In Table~\ref{tab:muc_onpeak} we show the reduction effects for the resonance production of the SM Higgs boson at 125 GeV for a muon collider (upper panel) including BES and ISR. 
The resonance production rate is reduced by a factor of 1.9 with the inclusion of ISR effect with the parameterization of Jadach-Ward-Was (b). Independently, the production rate would be reduced by factors of 4.2 and 1.7 for beam spread of 0.01\% and 0.003\% respectively.\footnote{In comparison with the cross sections considering beam energy spread in our initial study~\cite{Han:2012rb}, some small numerical differences are generated due to a different choice of the Higgs boson mass of 125 GeV instead of 126 GeV and correspondingly the different branching fractions and total widths.} The total reduction after the convolution of the beam spread and the ISR effect is 7.1 and 3.2 for the two beam spread scenarios, respectively. 

\begin{figure}[t]
\centering
\includegraphics[width=0.48\textwidth]{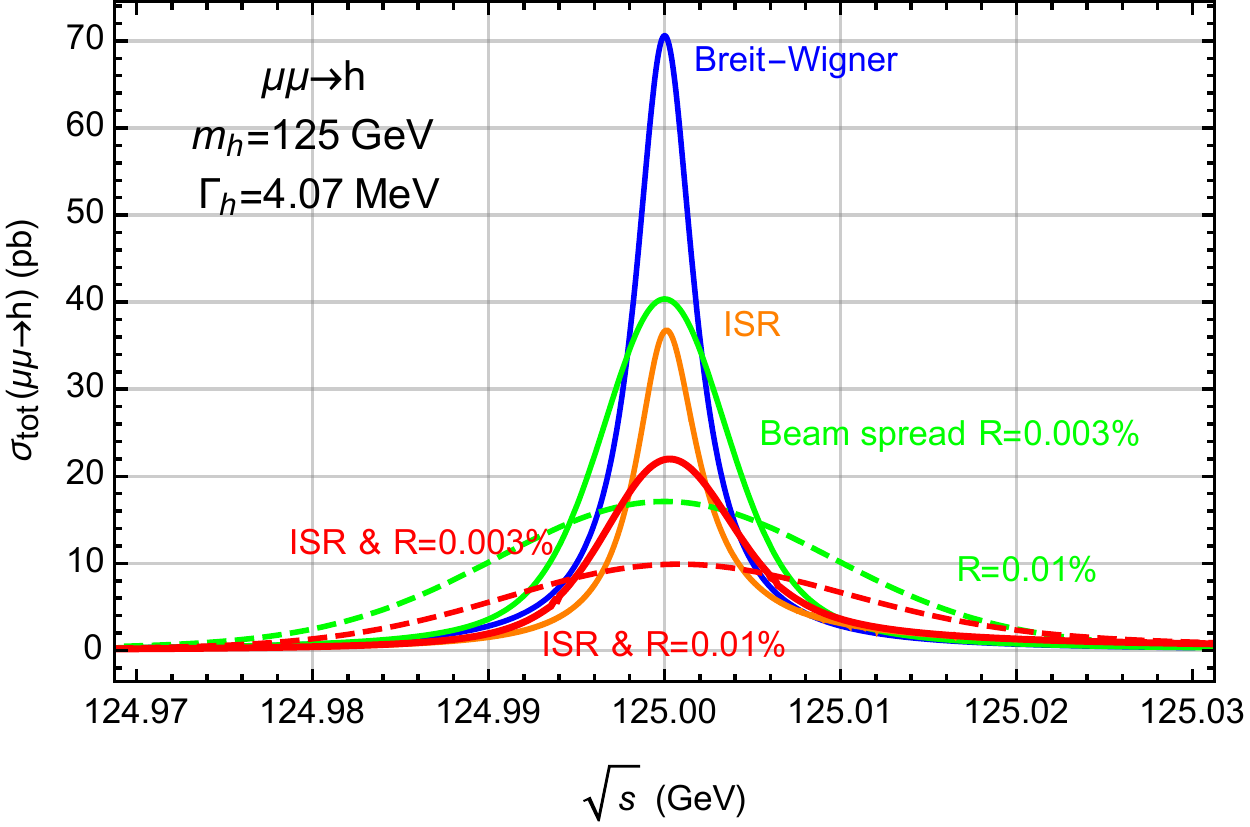} 
\includegraphics[width=0.48\textwidth]{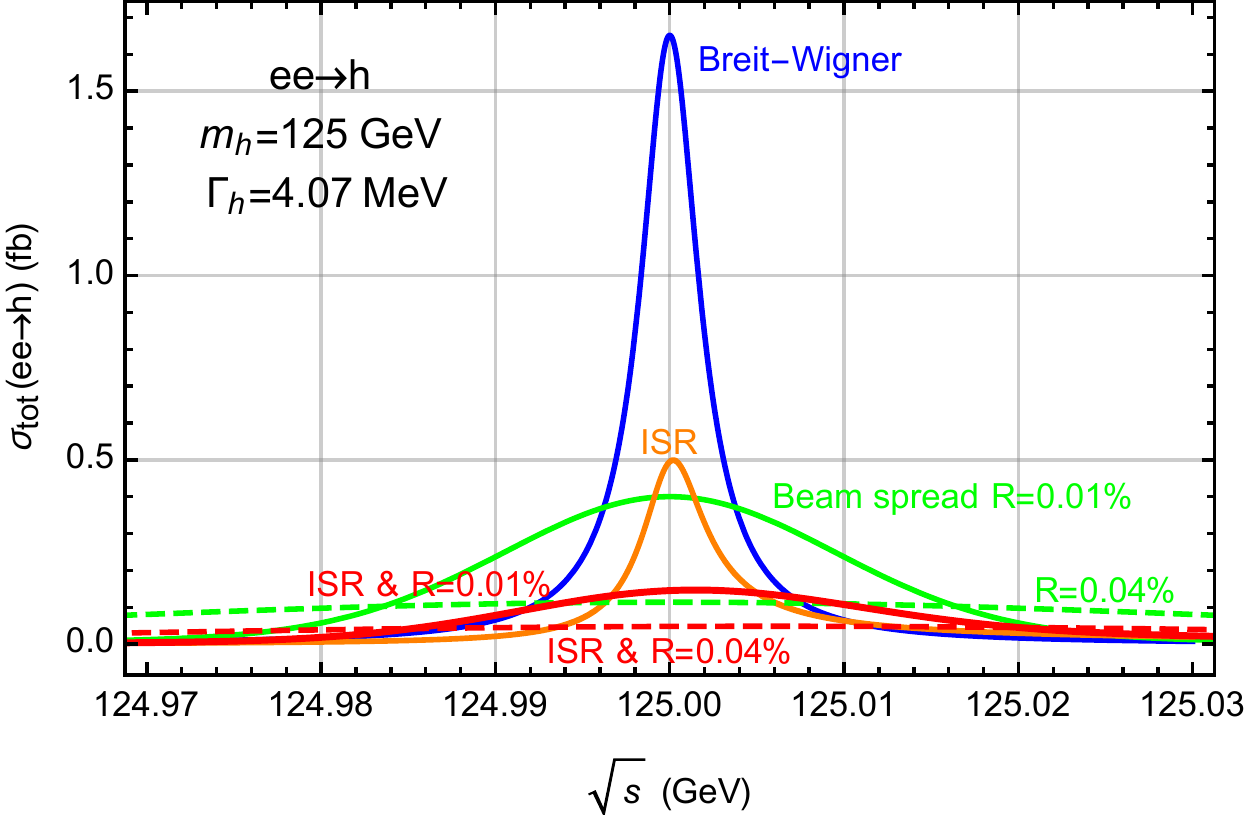}
\caption[]{The line shapes of the resonances production of the SM Higgs boson as a function of the beam energy $\sqrt s$ at a $\mm$ collider (left panel) and an $\ee$ collider (right panel). The blue curve is the Breit-Wigner resonance line shape. The orange line shape includes the ISR effect alone for Jadach-Ward-Was (b). The green curves include the BES only with two different energy spreads.
The red line shapes take into account all the Breit-Wigner resonance, ISR effect and BES 
in solid and dashed lines, respectively. 
}\label{fig:lineshape}
\end{figure}

To illustrate the resulting line-shape we show in Fig.~\ref{fig:lineshape} (left panel for a $\mm$ collider) for various setups of our evaluation. We show the sharp Breit-Wigner resonance in solid blue lines. The BES will broaden the resonance line-shape with a lower peak value and higher off-resonance cross sections, as illustrated by the green curves. The solid lines and dashed lines represent the narrow and wide BES of 0.01\% and 0.003\%, respectively. The ISR effect is asymmetric below and above the resonant mass, because it only reduces the collision energy by emitting photons, shown in the orange curve. In regions 10 MeV above the Higgs mass, the ISR effect increases the production rate via ``radiative return'' mechanism. Still, the overall effect is the reduction of on-shell rate as clearly indicated in the plot. In red lines we show the line shapes of the Higgs boson with both the BES and the ISR effect.  We can see the resulting line shape is not merely a product of two effect but rather complex convolution, justifying necessity of our numerical evaluation. 

\begin{table}[tb]
\centering
\begin{tabular}{|c|c|c|c|c|c|}
  \hline
 & $\mm\rightarrow h$ & \multicolumn{2}{|c|}{$h\rightarrow b\bar b$} &  \multicolumn{2}{|c|}{$h\rightarrow WW^*$}  \\ \cline{3-6}
 \raisebox{1.6ex}[0pt]{R (\%)} & $\sigma_{\rm eff}$ ({\bf pb}) & $\sigma_{Sig}$ & $\sigma_{Bkg}$ &  $\sigma_{Sig}$ & $\sigma_{Bkg}$ \\ \hline
 $0.01$ & $10$ & $5.6$ & & $2.1$ &  \\ \cline{1-3} \cline{5-5}
$0.003$ & $22$ & $12$ & \raisebox{1.6ex}[0pt]{$20$}  & $4.6$ & \raisebox{1.6ex}[0pt]{$0.051$}\\ \hline
  \hline
 & $\ee\rightarrow h$ & \multicolumn{2}{|c|}{$h\rightarrow b\bar b$} &  \multicolumn{2}{|c|}{$h\rightarrow WW^*$}  \\ \cline{3-6}
 \raisebox{1.6ex}[0pt]{R (\%)} & $\sigma_{\rm eff}$ ({\bf ab}) & $\sigma_{Sig}$ & S/B &  $\sigma_{Sig}$ & S/B \\ \hline
 $0.04$ & $48$ & $27$ & \multirow{2}{*}{$\mathcal{O}(10^{-6})$} & $10$ & \multirow{2}{*}{$\mathcal{O}(10^{-3})$} \\ \cline{1-3} \cline{5-5}
 $0.01$ & $150$ & $81$ &  & $31$ &  \\ \hline
\end{tabular}
\caption{Signal and background effective cross sections at the resonance $\sqrt {s}=\mh=125~\gev$ at a $\mm$ collider (upper panel, in pb) and an $\ee$ collider (lower panel, in ab) for two choices of beam energy resolutions $R$ and two leading decay channels with ISR effects taken into account, with the SM branching fractions $\br_{b\bar b}=58\%$ and $\br_{WW^*}=21\%$. For the $b\bar b$ background, a conservative cut on the $b\bar b$ invariant mass to be greater than 100 GeV is applied.
}
\label{tab:muon_sigbkg}
\end{table}

Having understood the ISR and BES effects on the signal production rates and line shapes, we now proceed to understand the effect on the background. For the muon collider study, the main search channels for the Higgs boson will be the exclusive mode of $b\bar b$ and $WW^*$.  For the $b\bar b$ final state the main background is from the off-shell $Z/\gamma$ $s$-channel production. The ISR and BES effects barely change the rate from such off-shell process. However, the ISR effect does increase the on-shell $Z\to b\bar b$ background through the ``radiative return'' mechanism.  Our numerical study shows that the ``radiative return'' of the $Z$ boson to $b\bar b$ increase the inclusive $b\bar b$ background by a factor of seven. Since we understand that the increase of the background is dominantly from the on-shell $Z$ boson, the new background rates after imposing a $b\bar b$ invariant mass cut of 95, 100, 110 GeV, change to 17, 20, 25 pb, respectively. Given the finite resolution of the $b$-jet energy reconstruction, we propose an invariant mass cut of the $b\bar b$ system of 100 GeV, which leads to around 20\% increase in such background comparing to the tree-level estimate.
\Edit{So far we have suggested the invariant mass cut for the $b\bar b$ pair, as an example of  discrimination from the background. One could also foresee a cut  on the angle between the two b-jets, which could be measured more precisely than the invariant mass.}\footnote{\Edit{We thank the Editor Gigi Rolandi for suggesting this discrimination procedure.}}

\begin{table}[tb]
\centering
\begin{tabular}{|c|c|c|c|c|}
  \hline
  $\Gamma_h=4.07~\mev$ & $L_{step}$ ($\fbi$) & $\delta{\Gamma_h}~(\mev)$ & $\delta B$ & $\delta m_{h}~(\mev)$ \\ \hline \hline
  \multirow{2}{*}{$R=0.01\%$} & 0.05 & 0.79 & 3.0\% &  0.36\\  \cline{2-5}
  & $0.2$ & $0.39$ & $1.1\%$ & $0.18$ \\ \hline \hline
  \multirow{2}{*}{$R=0.003\%$} &  0.05 & 0.30 & 2.5\% &  0.14\\ \cline{2-5}
  & $0.2$ & $0.14$ & $0.8\%$ & $0.07$ \\
  \hline
\end{tabular}
\caption{Fitting accuracies for one standard deviation of $\Gamma_h$, $B$ and $m_h$ of the SM Higgs with the scanning scheme for two representative luminosities per step and two benchmark beam energy spread parameters.}
\label{tab:acrcy}
\end{table}

Beyond the $b\bar b$ final state, another major channel for muon collider Higgs physics is the $WW^*$ channel. This channel enjoys little (irreducible) background form the SM process. The ISR effect introduces no ``radiative return'' for such process. Consequently, the background rate does not change from the tree-level estimate.  We summarize in table~\ref{tab:muon_sigbkg} the on-shell Higgs production rate and background rate in these two leading channels with the inclusion of the ISR and BES effects. We can see from the table that at the muon collider Higgs factory, the signal background ratio is pretty large and the observability is simply dominated by the statistics. The ``radiative return'' from the ISR effect, however, does impact several other Higgs decay channel search more. For example, searches of Higgs rare decay of $h\to Z\gamma$, Higgs decay of $h\to ZZ^*$ with $Z^*\to\nu\bar \nu$, etc are facing more challenges and new selection cuts need to be designed and applied. 


Finally, we perform a study on the potential precision on the Higgs properties at a future muon collider through a lineshape scan. We follow the benchmarks, statistical treatment and procedure defined in Ref.~\cite{Han:2012rb}, where a 21 steps scan in the mass window of $\pm 30$~MeV around the Higgs mass with equal integrated luminosities.\footnote{The Higgs mass may not known to the $\pm 30$~MeV level by the time of the muon collider, and a pre-scan stage to determine the Higgs mass will be required~\cite{Conway:2013lca}.} A fit to the result of such lineshape scan can simultaneously determine the Higgs total width $\Gamma_h$, the Higgs mass $m_h$ and interaction strength $B$ with great precision. The interaction strength $B$ can be directly translated into the Higgs muon Yukawa after fixing the decay branching fractions or performing a global fit. We tabulate the projected precisions on these quantities in Table.~\ref{tab:acrcy} for the two benchmark BES values of $R=0.01\%$ and $R=0.003\%$ and two benchmark integrated luminosities per scan step of $0.05\fbi$ and $0.2\fbi$. For the case of optimistic BES of $R=0.003\%$, we find that all related Higgs properties can be determined to great precision, including the Higgs width to 0.14~MeV, the Higgs mass to 0.07~MeV and the interaction strength $B$ to 0.8\% with $0.2~\fbi$ per scan step. For a lower statistics of $0.05~\fbi$ per scan step, the projected precision are basically doubled, following the statistical dominance argument. For the conservative BES of $R=0.01\%$ with different luminosities, the achievable precision is roughly the result for $R=0.003\%$ doubled, in more detail the precision on Higgs width and interaction strength are still great at percent level while the precision on Higgs mass remains at sub MeV level. 

\subsection{The case for the electron-positron collider}

The case for the electron-positron collider for a resonant production of the SM Higgs boson has a rather different physics purpose. Unlike the muon collider case for a precision measurement of many crucial properties of the Higgs boson, including the width, mass, muon Yukawa coupling with unprecedented precision, the most important physics goal of an electron-positron collider at 125 GeV is to constrain the electron Yukawa coupling. This would certainly be a first $O(1)\sim O(10)$ level test on this first generation Yukawa coupling. The potential to probe this first generation Yukawa will provide important implications for a broad class of flavor models. 

In electron-positron collisions, ISR effect is significantly larger because the radiation is inversely  proportional to lepton mass squared. The ISR effect is further amplified by the beamstrahlung due to the demand of a high instantaneous luminosity. These lead to a broadening of the beam energy distribution. The on-peak cross section is more notably reduced than that at a muon collider. 
In Table~\ref{tab:muc_onpeak} (lower panel), we show the on-resonance production rate reduction for the SM Higgs boson at 125 GeV. The on-resonance production rate is reduced by a factor of 3.4 with the inclusion of ISR effect. The achievable beam parameters for the possible electron-positron resonant Higgs factory is not clear so far. For the sake of demonstration, we choose two benchmark cases of the BES: $R = 0.04\%$, which is running design for the FCC-ee at Z-pole; and an improved design $R=0.01\%$ for possible future developments~\cite{FCCeebeam}. The production reduction factors are 14 and 4.2 for those two beam spreads, respectively. 
The total reduction after the convolution of the beam spread and the ISR effect is 35 and 11 for the two beam spread scenarios, respectively as shown in the last column in the table. 

To illustrate the resulting line-shape we show in Fig.~\ref{fig:lineshape} (right panel for an $\ee$ collider) for various setups of our evaluation. The sharp Breit-Wigner resonance is shown by the solid blue line. The BES will broaden the resonance line-shape with a lower peak value and higher off-resonance cross sections, as illustrated in the green curves, with the solid line and dashed line representing the two BES parameterizations of 0.04\% and 0.01\%, respectively. The resulting line-shape features are very similar to the case of a muon collider as shown in the left panel, but with larger reduction factors from both the BES and ISR effects. 

In Table~\ref{tab:muon_sigbkg}
at the lower panel, similar to the muon collider case, we list signal rates for the two leading decay channels of the SM Higgs boson for the electron-positron resonant Higgs factory. The signal rates for the two leading Higgs decay channels are all at tens of attobarn level. The background rates are the same as listed in the muon collider case in Table~\ref{tab:muon_sigbkg} and we hence list the signal background ratio $S/B$ instead. We observe that the $S/B$ for the $h\to b\bar b$ process is quite small $\mathcal{O}(10^{-6})$ and this channel will not be contributing much to the Higgs physics.  
Next, the $h\to WW^*$ will be the leading channel for the consideration, if assuming that the systematics can be controlled at $\mathcal{O} (10^{-3})$ level. 

\begin{figure}[t]
\centering
\subfigure{
\centering
\includegraphics[width=0.485\textwidth]{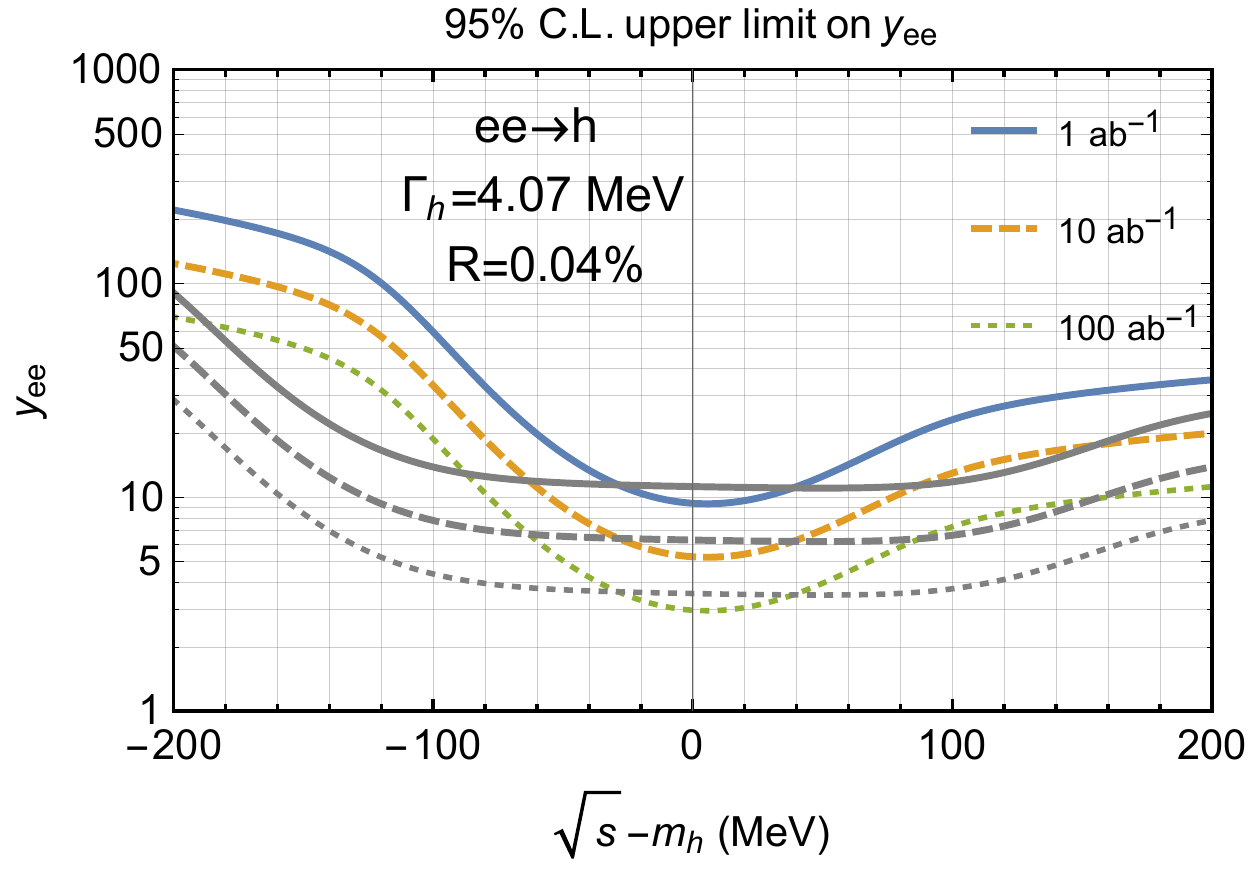}}
\subfigure{
\centering
\includegraphics[width=0.485\textwidth]{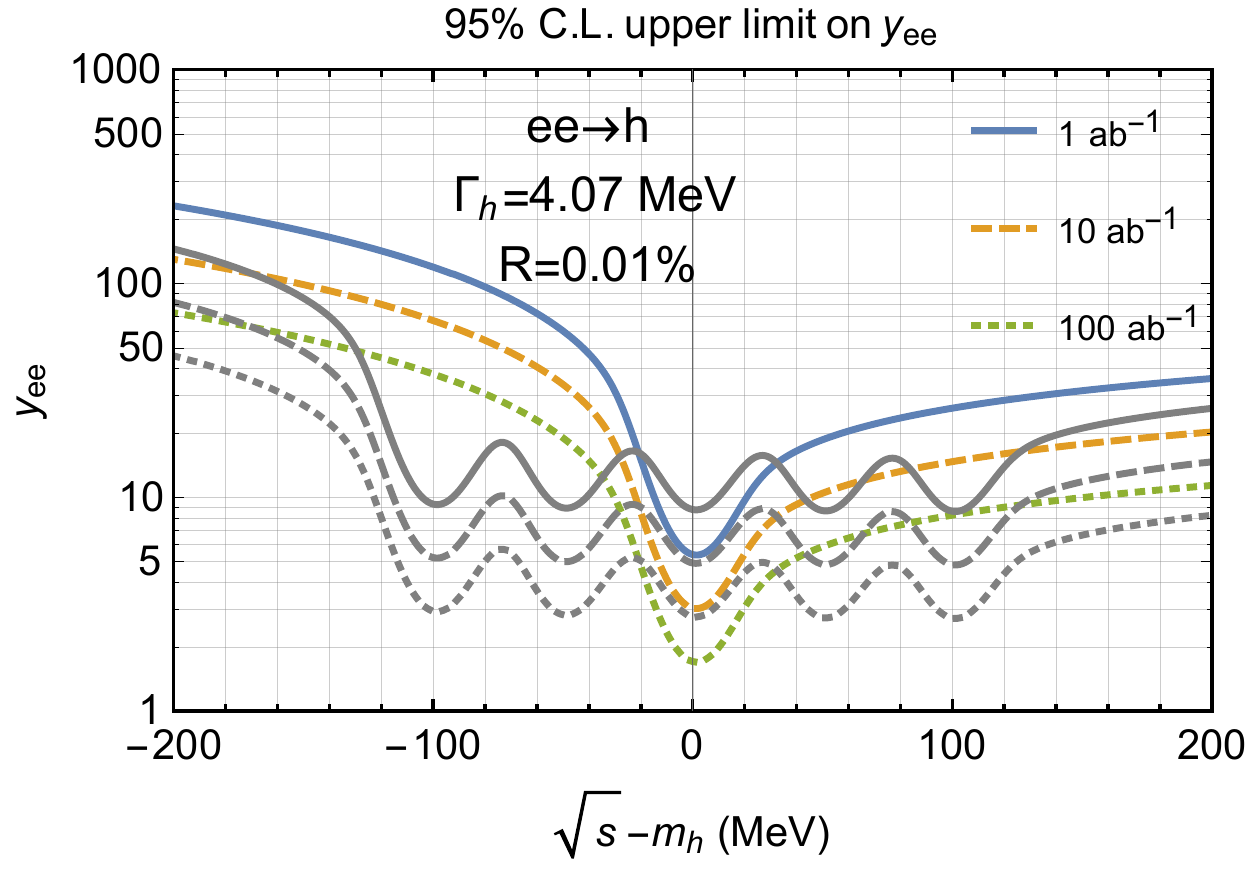}}
\caption[]{The projected 95\% C.L. upper limit on the Higgs electron Yukawa coupling at an electron-positron resonant Higgs factory with different integrated luminosities for two operational strategies. We consider two benchmark beam energy resolutions of 0.04\% and 0.01\% in the left panel and the right panel, respectively.
}\label{fig:eesensitivity} 
\end{figure}

In Fig.~\ref{fig:eesensitivity} we show the projected 95\% C.L. upper limit on the Higgs-electron Yukawa coupling (normalized to the SM value) with various collider running scenario and search strategies as a function of the difference between the beam energy and the Higgs pole mass ($\sqrt{s}-m_h$) based on the exclusive channel of $\ee\to h\to WW^*$.  We demonstrate two strategies here, one fixed energy for the full integrated luminosity (shown in colored lines) and one five-step scan with 50~MeV intervals around the Higgs mass (shown in gray lines) with equal shares of the total integrated luminosities.  The results for the two benchmark case of BES $R=0.04\%$ and $R=0.01\%$ are displayed in the left panel and right panel, respectively. In both panels, the exclusion limits are shown assuming null observation beyond the SM expectation, and from up to down, the solid, dashed and dotted lines represent the integrated luminosity of 1, 10, 100~$\abi$, respectively. 

As for the running strategy of sitting on a single energy for the total integrated luminosity, the upper limit on the electron Yukawa can reach 3, 5 and 8 times the SM value for $R=0.04\%$ (left panel) and 1.7, 3, 5 times the SM value for $R=0.01\%$ (right panel) with an integrated luminosity of 1, 10, 100~$\abi$, respectively, if the beam energy is tuned right at the Higgs mass. If the beam energy is set to be around 60~MeV (30~MeV) above the Higgs pole mass, the upper limit on the electron Yukawa is doubled. The asymmetric behavior of Higgs line-shape generated by the ISR effect appears here as the exclusion limits degrades much faster when the beam energy is below the Higgs mass than when above. However, we may not have {\it a priori} precise knowledge of the Higgs mass. 
We thus demonstrate an alternative strategy of a 5 step scanning with an interval of 50~MeV around the Higgs mass in gray lines. We can see in the gray curves that this strategy provides a relative flat sensitivity across the $\pm 100$~MeV range for BES $R=0.04\%$, yielding an upper bound of around $3.5\sim 4$ times the SM electron Yukawa with 100~$\abi$ integrated luminosity. For BES $R=0.01\%$ The wavy structure in the exclusion limits indicates the divide of the scanning steps is not fine enough, and the exclusion at 100~$\abi$ various between 3 to 5 times the SM electron Yukawa. We conclude that for a single energy run, the better beam energy resolution than $R=0.04\%$ is not advantageous unless a knowledge on the Higgs mass precision of around 10~MeV is available before choosing the beam energy. If the Higgs mass is known to a level of $\pm 50$~MeV, a multistep scan can provide a rather uniform exclusion limits in the $\pm 100$~MeV window of the Higgs mass, reaching around 3 times the SM electron Yukawa, and in this case better BES simply means more scanning steps in this mass window.

\section{Conclusion}

We studied the effects from the initial state radiation and beam energy spread coherently for lepton colliders for the narrow Higgs boson production in the $s$-channel.
We presented a few representative choices of the BES and different approximations for the ISR. 
%
We quantify their impact in different scenarios for both the Higgs signal and SM background. 
We found that
\begin{itemize}
\item The BES effect is potentially the leading factor for the resonant signal identification, and it alone reduces the on resonance Higgs production cross section by a factor of 1.7 (4.2) for a muon collider with $R=0.003\%$ ($R=0.01\%$), and by a factor of 4.2 (14) for an electron-positron collider with $R=0.01\%$ ($R=0.04\%$), as shown in Table \ref{tab:muc_onpeak}.
\item  The ISR effect alone reduces the on\Edit{-}resonance Higgs production cross section by a factor of 1.9 for a muon collider and 3.4 for a electron-positron collider (Table \ref{tab:muc_onpeak}). The ISR effect is asymmetric above and below the Higgs pole mass, and slightly shift the location of the peak cross section, as shown in Fig.~\ref{fig:lineshape}.
\item  The total reduction factors for the on\Edit{-}resonance Higgs production cross section after convoluting the BES and ISR effects are 3.2 (7.1) for a muon collider with $R=0.003\%$ ($R=0.01\%$), and 11 (35) for a electron-positron collider with $R=0.01\%$ ($R=0.04\%$), as tabulated in the last column in Table~\ref{tab:muc_onpeak}.
\item The background for the $h \to b\bar b$ channel is increased by a factor of  seven due to the ``radiative return'' of the $Z$ boson at lepton colliders and a cut on the minimal $b\bar b$ invariant mass of 100 GeV reduces such background, resulting in an increase of the tree-level estimate of the background by 20\%. For a muon collider, both the $h\to b\bar b$ and $h\to WW^*$ contribute to the signal sensitivity. For an electron-positron collider, only the $WW^*$ contributes due to the smallness of $S/B$ for the $h\to b\bar b$ channel.
\item For a muon collider resonant Higgs factory with our more robust study including both the BES and ISR effects, a 21 steps scan in the $\pm 30$~MeV window around the Higgs mass would provide percent level precision on the Higgs width and Higgs muon Yukawa coupling measurements, and sub MeV precision on the Higgs mass determination for various collider configurations, as tabulated in Table \ref{tab:acrcy}.
\item  For an electron-positron Higgs resonance factory, since a pre-scan to determine the precise Higgs mass is not feasible, one can achieve an 95\% C.L. upper limit of $3\sim 5$ ($5\sim 10$) times the SM electron Yukawa by scanning through the $\pm 100$~MeV window around Higgs mass with $100~(10)~\abi$ integrated luminosity, as shown in Fig.~\ref{fig:eesensitivity}.
\item  For an electron-positron Higgs resonance factory, increasing beam quality by reducing the beam energy spread does not increase the sensitivity to the Higgs electron Yukawa coupling, as {\it a priori} knowledge of the precise Higgs mass better than $10$~MeV may not be available. Hence a first run at 240$-$250~GeV mode for an electron-positron collider maybe a step to make better result out of such resonance Higgs factory. A muon collider Higgs factory, on the other hand, is very complimentary and can provide sub~MeV level of Higgs mass determination, which optimizes the sensitivity for the potential electron-positron Higgs resonance factory provided great beam quality can be achieved. In this case, a single run at fixed $\ee$ energy can achieve an upper limit of 1.7 (3) times the SM electron Yukawa for BES $R=0.01\%$ with $100~(10)~\abi$ integrated luminosity, as shown in Fig.~\ref{fig:eesensitivity}.
\end{itemize}

Our study improves the existing analyses of the proposed future resonant Higgs factories and provides further guidance for the target accelerator designs with respect to the physical goals.

\begin{acknowledgments}
This work is supported in part by the U.S.~Department of Energy under grant No. DE-FG02-95ER40896, in part by PITT PACC. Fermilab is operated by Fermi Research Alliance, LLC under Contract No. DE-AC02-07CH11359 with the U.S. Department of Energy. Z.L. and T.H. thank the Kavli Institutes for Theoretical Physics at UC Santa Barbara and in China at the CAS, respectively, for their hospitality during the final stage of this paper.
\end{acknowledgments}

\appendix

\section{Analytical formulas for the initial state radiation}
\label{app:ISR}

We list a few commonly used analytical formulas for the ISR. We first introduce $\beta_\ell$ as a common loop factor for radiation effects
\beq
\beta_\ell=\frac {2\alpha} \pi \left( \log\frac {\hat s} {m_\ell^2}-1\right),
\eeq
where $\alpha$ is the fine-structure constant evaluated at the collision center of mass energy $\sqrt {\hat s}$ and $m_\ell$ is the charged lepton mass. The difference between the electron and muon is mainly carried in this factor.
We present the ISR effect for various parameterization and expansion order in the following, where we annotate the dependence on charged leptons explicitly with the subscript~$\ell$.
{\flushleft
$\bullet$~Kuraev-Fadin~\cite{Kuraev:1985hb}:}
\begin{flalign}
&f_{\ell\ell}^{\rm ISR; KF} (x; \hat s)=\int_{x}^1 dy~2 f_{\ell}^{\rm KF} (y; \hat s)f_{\ell}^{\rm KF} (\frac x y; \hat s),&
\label{eq:KF}\\
&{\rm with~}f_{\ell}^{\rm KF} (x; \hat s)=\frac {\beta_\ell} {16} \left((8+3\beta_\ell)(1-x)^{\frac {\beta_\ell} 2-1}-4(1+x)\right).&\nonumber
\end{flalign} 
{\flushleft
$\bullet$~Nicrosini-Trentadue~\cite{Nicrosini:1986sm}:}
\begin{flalign}
&f_{\ell\ell}^{\rm ISR; NT} (x; \hat s)=\\
&\Delta_\ell \beta_\ell x^{\beta_\ell -1}-\frac 1 2 \beta_\ell  (2-x)+\frac 1 8 \beta_\ell ^2\left((2-x)(3\log(1-x)-4\log x)-\frac {4\log(1-x)} {x} -6+x\right)+\mathcal{O}(\beta_\ell ^3)\nonumber \\
&{\rm with}~~~\Delta_\ell=1+\frac {\alpha} {\pi} (\frac 3 2 \log\frac {s} {m_\ell^2}-\frac 1 3 \pi^2 -2)+\nonumber\\
&\frac {\alpha^2} {\pi^2} \left(\left(\frac 9 8 - 2 \zeta[2] \right)\log \frac {\hat s} {m_\ell^2}+\left(-\frac {45} {16}+\frac {11} {2}\zeta[2]+3\zeta[3]\right)\log \frac {\hat s} {m_\ell^2}-\frac 6 5 \zeta^2[2]-6\zeta[2]\log 2+\frac 3 8 \zeta[2]+\frac {57} {12} \right),\nonumber
\label{eq:NT}
\end{flalign}
where $\zeta[n]$ is the Euler-Riemann zeta function.
{\flushleft
$\bullet$~Jadach-Ward-Was~\cite{Jadach:2000ir,Jadach:2015cwa}:}
\begin{flalign}
&f_{\ell\ell}^{\rm ISR; JWW(a)} (x; \hat s)=e^{\frac {\beta_\ell} 4 + \frac \alpha \pi \left(-\frac 1 2 + \frac {\pi^2} 3\right)}\frac {e^{-\gamma\beta_\ell}} {\Gamma[1+\beta_\ell]}\beta_\ell (1-x)^{\beta_\ell-1}\\
&f_{\ell\ell}^{\rm ISR; JWW(b)} (x; \hat s)=f_{\ell\ell}^{\rm ISR; JWW(a)} (x; \hat s)\left(1+\frac {\beta_\ell} 2  - \frac 1 2 (1-x^2)\right) \nonumber\\
&f_{\ell\ell}^{\rm ISR; JWW(c)} (x; \hat s)=f_{\ell\ell}^{\rm ISR; JWW(a)} (x; \hat s)\left(1+\frac {\beta_\ell} 2 +\frac {\beta_\ell^2} 8 - \frac 1 2 (1-x^2)+\beta_\ell\left(-\frac {1-x} 2-\frac {1+3 x^2} 8 \log x\right)\right),\nonumber
\label{eq:JWW}
\end{flalign}
where $\gamma$ is the Euler-Mascheroni constant. Their numerical results are compared  in Fig.~\ref{fig:ISRpdf} and the corresponding discussion in the text. 

\bibliographystyle{JHEP}
\bibliography{references}

\end{document}